\documentclass[a4paper]{jpconf}
\usepackage{graphicx}
\begin{document}
\title{A search for optical counterparts of the complex Vela~X system}

\author{T E Marubini$^1$, R R Sefako$^2$, C Venter$^1$, and O C de Jager$^1$ }

\address{$^1$North-West University, Potchefstroom Campus, Private Bag X6001, Potchefstroom, 2520, South Africa\\
$^2$South African Astronomical Observatory, P.O. Box 25, Sutherland, 6920, South Africa}

\ead{21610126@nwu.ac.za}

\begin{abstract}
The pulsar wind nebula (PWN) associated with the Vela pulsar is a bright source in the radio, X-ray and gamma-ray bands, but not in the optical. This source is very near, lying at a distance 
of 290 pc, as inferred from the radio and optical parallax measurements of the pulsar. Knowledge of the brightness and structure of the Vela PWN in optical is important in order to constrain the underlying particle spectrum (and possibly the $B$-field properties and particle losses) associated with this extended source. We use results from the Digital Sky Survey, as well as results obtained using the SAAO 1.0 m telescope equipped with an imaging CCD (STE4) and BV filters, in an attempt to measure optical radiation from Vela~X. To enlarge our field of view, we constructed a mosaic  consisting of 3 $\times$ 3 frames around the pulsar position. We present spectral measurements from the {\it High Energy Stereoscopic System (H.E.S.S.)}, \textit{Fermi Large Area Telescope (LAT)}, {\it ASCA}, \textit{Hubble Space Telescope (HST)}, \textit{Very Large Telescope (VLT)}, \textit{New Technology Telescope (NTT)}, and \textit{Wilkinson Microwave Anisotropy Probe (WMAP)}, in addition to our  optical results. Using these data,
 we investigate whether or not the radio synchrotron component can be smoothly extrapolated to the optical band. This would constrain the electron population to consist of either a single or multiple components, representing a significant advancement in our understanding of this complex multiwavelength source. 
\end{abstract}

\section{Introduction}
Vela~X has been a very enigmatic source since its discovery, and its true nature has been the subject of debate for many years before its verification as a Crab-like pulsar wind nebula (PWN; see Section~\ref{sec:disc}). It is a very rich source, and has been studied extensively in all wavebands. The complex system includes the PWN (Section~\ref{sec:PWN}), Vela supernova remnant (SNR; Section~\ref{sec:SNR}), and the well-known Vela pulsar (Section~\ref{sec:pulsar}), with each of the wavebands giving a unique window on the intricate properties of this source. 

In this paper, we describe our efforts to detect optical emission from the Vela~X plerion. This is a crucial piece of the puzzle needed to constrain important properties, such as the leptonic injection spectrum, $B$-field, or diffusion coefficient, all of which impact on the broadband spectral emission of Vela~X. 

In Section~\ref{sec:Results} we describe our data accumulation, analysis and results, while we discuss our conclusions in Section~\ref{sec:Conclusion}.

\subsection{Discovery and identification}
\label{sec:disc}
Since the detection of Vela~X, a disagreement raged between various authors as to the true nature of this source. Three strong nonthermal radio sources of enhanced brightness temperature have been observed~\cite{Rishbeth58} and designated Vela~X, Vela Y, and Vela Z. To determine the nature of  these sources several authors have tried to find their spectral indices \cite{Milne68, Milne80, Weiler80, Milne86, Dwarakanath91}. A controversy among two groups (\cite{Milne68, Milne86} vs.\ \cite{Weiler80}) erupted. The first group claimed that there was no significant difference between the indices of  Vela~X and Vela YZ, and concluded that Vela~X was associated with the Vela supernova remnant  (SNR). Conversely, the second group maintained that Vela YZ exhibited a steeper index than that of  Vela~X, and concluded that Vela YZ is part of the shell SNR, while Vela~X was a plerion (defined as a Crab-like PWN, a filled-centre SNR). To solve this mystery, the spectral indices were remeasured~\cite{Dwarakanath91} and the 
free-free absorption at 85~MHz estimated. The plerion classification was thus confirmed. Vela~X was furthermore found to consist of a network of fine,  overlapping, linear, synchrotron-emitting filaments ubiquitous in other plerions~\cite{Frail97}.  The plerion identification was clinched by showing that there are significant morphological differences between the nebula and the shell filaments~\cite{Bietenholz91}. Spectral indices for Vela~X, Vela~Y, and Vela~Z have been obtained~\cite{Alvarez2001} which were consistent with a spherical SNR containing a central plerion. We therefore conclude that Vela~X is a plerion.

\subsection{Multiwavelength properties of the Vela PWN}
\label{sec:PWN}
Two radio lobes oriented about the spin axis of the pulsar between the energy 1.4~GHz$-$8.5~GHz have been uncovered~\cite{Lewis02}, in agreement with models where the emission is driven by particles from its magnetic poles. They also found that the northern lobe exhibited a bright edge, while the southern  lobe was more diffuse. Radio emission from two lobes around the pulsar between the energy 2.4~GHz$-$5~GHz was next measured~\cite{Dodson2003}, the lobes differing in size and brightness, with the radio emission starting where the X-ray emission stops. The orientation of the 31~GHz PWN is in agreement with that of the radio lobes observed at higher resolution and at lower frequencies~\cite{Hales04}. 

A marginal detection of an optical counterpart to the compact Vela~X-ray nebula was claimed~\cite{Oegelman89}. Much deeper upper limits of 28.1 \textit{HST} mag arcsec$^{-2}$ for the inner nebula, and $28.0-28.5$ \textit{HST} mag arcsec$^{-2}$ for the outer nebula have subsequently been found~\cite{Mignani2003}. An optical plerionic excess for Vela has been observed~\cite{Sefako02}, as well as an H$\alpha$ feature at the bow shock of Vela, and also an apparent excess around the pulsar position in R band. 

The X-ray source 2U 0832.45 was detected above 9$\sigma$~\cite{Kellogg73}, being less than 1$^\circ$.2 in extent. Various authors found spectral indices ranging from $\Gamma = 1.7- 2.17$ for different regions of the Vela~X plerion. A compact X-ray nebula with extension $\sim1^{\prime}$ was measured~\cite{Oegelman89}, as well as an emission region extending from the pulsar in the southwestern  direction out to $3^{\prime}$. An X-ray emitting ``jet'' (cocoon)  originating at the pulsar, with a size of $45^{\prime}\times~12^{\prime}$, extending towards the  south-southwest was next observed~\cite{Markwardt95}. The cocoon's X-ray spectrum was similar to the  spectrum of the surrounding Vela SNR~\cite{Markwardt97}. Displacement and flux changes in various jet substructures have been observed~\cite{Pavlov01}, as well as a hint of a counterjet with a position angle of 130$^\circ$, measured east of north, and aligned to within $8{^\circ}\pm5^{\circ}$ with the proper motion vector. Two arcs of emission around the 
pulsar, surrounded by diffuse emission in a ``kidney-bean shape'' were seen~\cite{Helfand01}. A two-component X-ray spectrum consisting of a thermal and a nonthermal component have been inferred~\cite{LaMassa08}. The spectrum was steepening with distance, rather than softening, which may be caused by the crushing of the cocoon against the PWN, leading  to adiabatic heating that hardens the spectrum. 

An unpulsed power-law spectrum with a steep photon index of $\Gamma = 1.6\pm 0.5$ in the energy range between 0.06~MeV and 0.4~MeV was obtained~\cite{Dejager_nebula,Dejager96,Strickman96}. These results are consistent with an extrapolation of the $\sim$ E$^{-1.7}$ spectrum of the $1^{\prime}$ compact nebula. An index of $\Gamma=2.2$ below 20~keV, with a significant steepening above $\sim$20~keV ($\Gamma= 2.7$), and a hardening above $\sim60$~keV to meet the index of $\Gamma= 1.8$ for the \textit{OSSE} spectrum (corresponding to the 1$^{\prime}$ compact nebula) have been obtained~\cite{Strickman96}. The Vela~X region to the south of the Vela pulsar has been observed with the High Energy Stereoscopic System (H.E.S.S.)~\cite{Aharonian2006}, yielding a fit spectrum between 550~GeV and 65~TeV with a photon index of $\Gamma= 1.45\pm0.09_{\rm stat}\pm0.2_{\rm sys}$ and an exponential cutoff energy of $13.8\pm2.3_{\rm stat}\pm4.1_{\rm sys}$ TeV for an integration region of 0$^{\circ}$.8. The best-fit intrinsic 
width of the very high energy (VHE) source was 0$^{\circ}.48\times0^{\circ}$.36 (i.e., $58\times$43 arcmin$^{2}$ vs.\ the X-ray size of $42\times12$ arcmin$^{2}$). \textit{AGILE} detected a source in the off-pulse window of the Vela pulsar at $\sim5.9\sigma$ significance, with a flux of ($3.5\pm0.7$) $\times10^{-7}$ cm$^{-2}$ s$^{-1}$ above 100 MeV, located $\sim 0^{\circ}.5$ southwest of the pulsar and having an extent of $1^{\circ}.5\times 1^{\circ}$~\cite{Pellizzoni10}. The \textit{Fermi} Large Area Telescope (LAT) detected $\gamma$-ray emission at the 14$\sigma$ level from a $2^{\circ} \times 3^{\circ}$ area south of the Vela pulsar~\cite{Abdo2010}. They found that the flux is significantly spatially extended, with a best-fit radius of $0^{\circ}.88\pm0^{\circ}.12$ for a uniform disc morphology. The spectrum is well described by a power-law with a spectral index of $2.41 \pm 0.09 \pm 0.15$.

\subsection{Multiwavelength properties of the Vela supernova remnant (SNR)}
\label{sec:SNR}
The total-power radio image of Vela exhibits many filaments and loop-like structures, and the northern side of the SNR displays a pair of bright concentric arcs, defining the northern edge of the shell~\cite{Duncan96}. On the southern edge, the emission fades smoothly into the background. The centre of the X-ray shell coincides with the centre of the optical SNR~\cite{Miller73}. An optical region covered by the Vela SNR filaments was found~\cite{Bergh73} having a diameter of 270$^{\prime}$. These filaments occur more densely near the region of Vela~X and do not appear to correspond to the radio hot spots. The Vela SNR is interacting with a corner of the observed HI shell at its northwestern border, and it was suggested that this interacting may be responsible for various local bright ultraviolet (UV) and optical arches and filaments, and also for the off-centre position of the Vela pulsar~\cite{Dubner92}. Far ultraviolet (FUV) emission was detected from the Vela SNR, subtending $\sim8^{\circ}$~\cite{
Nishikida2006}. A faint X-ray shell was seen~\cite{Aschenbach95}, which is described by a circle of $\sim8^{\circ}$.3 diameter, showing that the pulsar is $\sim25^{\prime}$ from the centre of the SNR. These authors also observed six extended X-ray features located outside the blast wave front having conal shapes. 

\subsection{Multiwavelength properties of the Vela pulsar}
\label{sec:pulsar}
The Vela pulsar is a rotation-powered neutron star (NS). Its discovery~\cite{Large68} revealed a period of $P = 89$~ms and a period derivative of $\dot{P} = 1.25\times10^{-13}$~s~s$^{-1}$, which implies an age of $\sim$11~000 yrs and a spin-down luminosity of $\dot{E} = 7\times10^{36}$~erg~s$^{-1}$~\cite{LaMassa08}. The Vela pulsar emission
in all wavelengths is less than $2\times10^{-4}$ that of the Crab~\cite{Wallace77}. 
A pointlike object in infrared (IR), the likely counterpart of the Vela pulsar, was found~\cite{Danilenko11}, and it was noted that the multiwavelength spectrum of the Vela pulsar reveals a steep flux increase towards the IR. 
A pointlike object coinciding with the pulsar in the energy range 0.1~keV to 4.5~keV has been found~\cite{Harnden85}. Thermal X-ray radiation was seen from the vicinity of the Vela pulsar~\cite{Oegelman89_A}. A power-law with spectral index of $\Gamma = 1.5 \pm 0.3$ for the soft component, and $\Gamma = 2.7 \pm 0.4$ for the harder component between the energy range 0.2~keV to 8.0~keV was found~\cite{Pavlov01}. A spectrum described by a power-law with a spectral index of $1.89\pm 0.06$ in the energy range between 50~MeV and 3~GeV was obtained~\cite{Kanbach80}, and they concluded that there was no evidence of a steady $\gamma$-ray emission from the direction of the pulsar. An average flux of ($7.8\times1.0$)$\times10^{-6}$ photons~cm$^{-2}$~s$^{-1}$ was detected~\cite{Kanbach94} and obtained a power-law spectrum with index $\Gamma = -1.70\pm0.02$ (30 MeV to 2 GeV). \textit{Fermi} found a phase-averaged power-law index of $\gamma=1.51^{+0.05}_{-0.04}$, with an exponential cutoff at $E_{\rm c} = 2.9 \pm 0.
1$~GeV~\cite{Abdo09}. An upper limit to the pulsed emission from the pulsar of ($3.7\pm0.7$)$\times10^{-13}$~photons~cm$^{-2}$~s$^{-1}$ was found above 2.5 TeV~\cite{Yoshikoshi97}. 

\section{Data analysis and results} 
\label{sec:Results}

\begin{figure}
\begin{center}
\includegraphics[width=13cm]{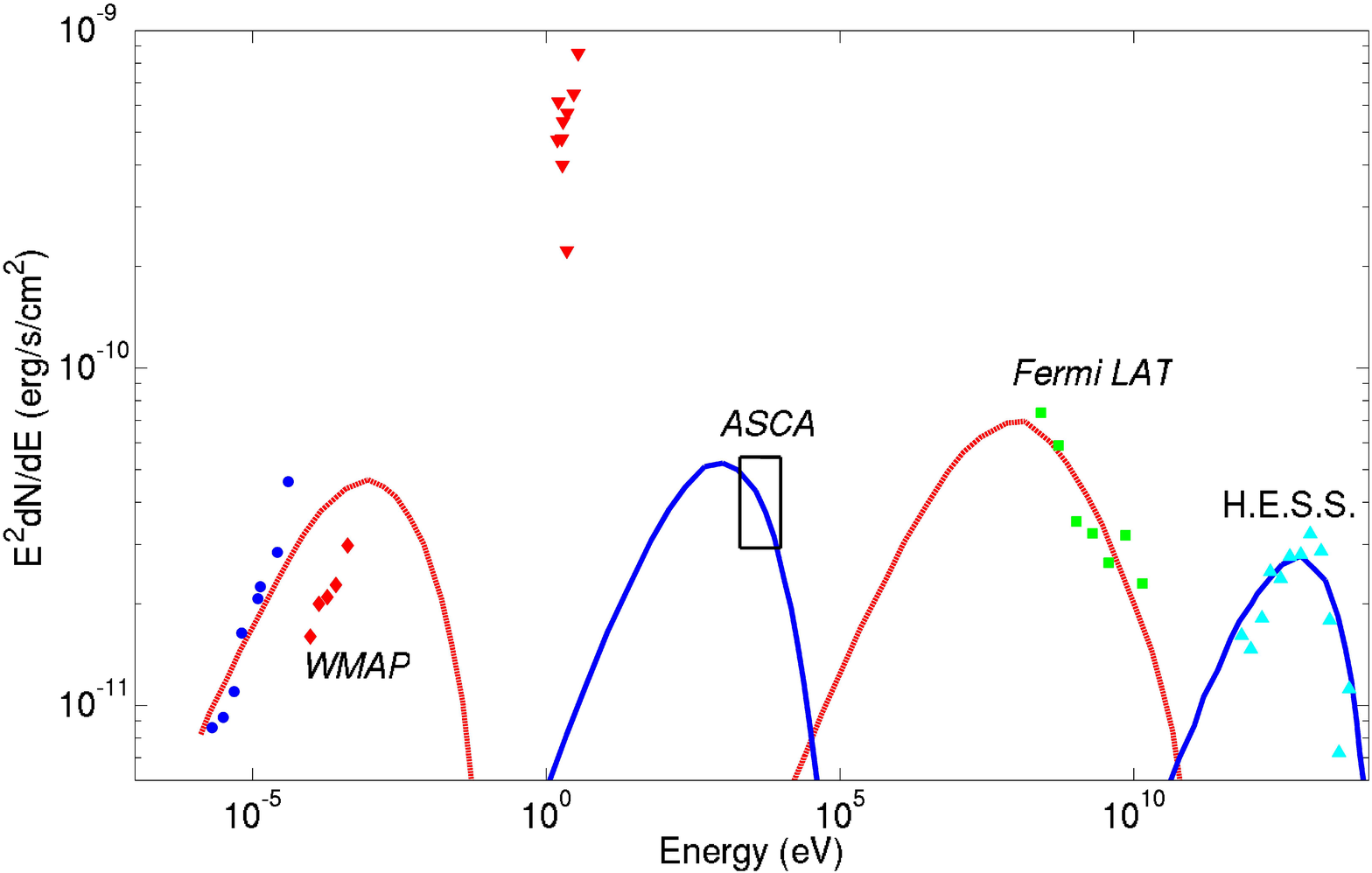}
\end{center}
\caption{\label{fig:VelaX_SED} Spectral energy density plot of Vela~X. Data and models are taken from~\cite{Hinton11}. The optical upper limits (red triangles) have been calculated using~\cite{Mignani2003}. The red dashed lines are for the extended radio nebula (ERN) while the blue lines are for the cocoon.}
\end{figure}

We have used the 1.0~m telescope at SAAO and collected data during February 2010 and March 2011. We took nine $5.3^\prime\times5.3^\prime$ images about the Vela pulsar, and constructed a mosaic consisting of $3\times3$ frames. The raw data of the illuminated detector of a $512 \times 512$ pixel chip (STE) were trimmed to $510 \times 510$ to avoid gaps in the data. An overlap from frame to frame of about $1^\prime$ was allowed and averages were taken at positions where overlapping occurred. We next rebinned our data to obtain a final trimmed image consisting of 70 rows and 72 columns, taking averages of smaller bins making up the new larger bins so as to decrease fluctuations in our data.

First, we attempted to resolve optical structure in our image mosaic. We removed the bright (mostly foreground) stars by hand, replacing them by the average background level. Different regions around the pulsar have slightly different backgrounds. The emission levels of a number of pixels with a very large number of counts were also replaced with the local background level to smooth out the image. Some excess emission is seen near the southern radio lobe position. This is suspected to be as a result of bright stars that were not clearly removed, but could also be the hints of the optical counterparts of Vela~X. 

We next proceeded to obtain upper limits. We constructed a circular source centred on the pulsar position, and by eye limited its size so as to exclude most of the obvious foreground stars. This yielded a source encompassing about 340~arcsec$^2$. From the number of counts $DN$ included by this source circle during our $\tau=120$~s long observation, we found an upper limit on the magnitude using~\footnote{www.stsci.edu/instruments/wfpc2/Wfpc2\_dhb/wfpc2\_ch52.html}
\begin{equation}
 m^\prime = -2.5\log_{10}\left(\frac{DN}{\tau}\right) + m_0, 
\end{equation}
with $m_0$ the zero point magnitudes of the telescope. The zero point magnitudes of the 1.0 m SAAO telescope are 22.0 for the {\it B} band, and 22.4 for the {\it V} band. Using standard stars, we next corrected our magnitudes as follows
\begin{equation}
 m = 1.03m^\prime-0.8. 
\end{equation}
Finally, we calculated the implied flux using
\begin{equation}
 F_\nu = F_{\nu,0}10^{-0.4m}. 
\end{equation}
We used $F_{\nu,0}=4~260$~Jy for the {\it B} filter, and $F_{\nu,0}=3~640$~Jy for the {\it V} filter. To derive an upper limit for the full cocoon size, we scaled our flux to extend over an area of $45\times12$~arcmin$^2$, making the implicit assumption that the background is roughly similar for the region centred on the pulsar and the cocoon region. Our $\nu F_{\nu}$ upper limits are $\sim9.6\times10^{-8}$~erg\,s$^{-1}$\,cm$^{-2}$ for the {\it B} filter, and $\sim7.5\times10^{-8}$~erg\,s$^{-1}$\,cm$^{-2}$ for the {\it V} filter.

In order to compare with other optical studies, we converted {\it U, B, V, R,} and {\it I} upper limits found~\cite{Mignani2003} using {\it Hubble Space Telescope (HST)}, {\it New Technology Telescope (NTT)}, and {\it Very Large Telescope (VLT)} to $\nu F_{\nu}$ values assuming the optical source size to be the same as that of the X-ray cocoon. These upper limits are indicated as red triangles in Figure~\ref{fig:VelaX_SED}, along with multiwavelength data and models for the broadband spectrum of Vela~X~\cite{Hinton11}.

\section{Conclusions} 
\label{sec:Conclusion} 
The search for the Vela~X PWN in optical is crucial to understand its morphology and how energy is distributed throughout the multiwavelength spectrum. It is a challenge to resolve the extended structure of Vela~X in the optical waveband, given that many sources contribute to the background radiation, as well as its apparent intrinsic faintness. 

We only observed a background level of optical emission near the Vela pulsar position, and there is no structure visible that could be associated with a possible Vela~X optical counterpart, apart from some hints of brighter emission near the position of the southern radio lobe. However, this may be due to bright stars that were not cleanly removed. Our $\nu F_{\nu}$ upper limits of $\sim10^{-7}$~erg\,s$^{-1}$\,cm$^{-2}$ are clearly not constraining for a multiwavelength model of Vela~X's emission. 

We note that our upper limits suffer from several observational factors that introduce uncertainties. Removal of bright or even fainter unresolved point sources remain problematic. Our integration times were also very short. Furthermore, the field of view of the telescope is too small to capture the full extension of the source, and we had to use a method of mosaicing several single images to obtain one large image that encapsulates the probable extension of the optical source. Changing conditions between the constituent frames add complexity and uncertainty to our analysis.

Neither our optical limits nor those of others prove to be constraining for the spectral model of Vela~X. Barring the gamma-ray data for the moment, one may model the radio, optical and X-ray data with either a single or double synchrotron component, implying either a single or multi-component electron spectrum interacting with the PWN $B$-field. However, the {\it Fermi} and H.E.S.S.\ data are consistent with a two-component leptonic interpretation~\cite{Abdo2010}, first put forward by~\cite{deJager07} and subsequently elaborated on by~\cite{deJager08}. A second model~\cite{Hinton11} invokes a single electron population, but then assumes different diffusive properties for the $2^\circ\times3^\circ$ extended radio nebula (ERN) and $45^\prime\times12^\prime$ cocoon regions (although the H.E.S.S.\ ``coccoon'' is quite more extensive). The particles are assumed to be contained within the ERN until the reverse shock from the SNR interacts with the ERN, and diffusive transport starts to take place. Older 
particles escape then from the ERN, while newly-injected particles are thought to suffer significant cooling in the high-$B$-field region near the pulsar before being injected into the low-$B$-field cocoon. Within such a scenario, each substructure (the ERN and cocoon) is responsible for different components of the multiwavelength spectrum (radio and GeV vs.\ X-ray and TeV). In essence, this is a different way of constructing a two-component electron injection spectrum. Our optical limits are not violated by these models, although the available limits support the idea of particle escape from the ERN (i.e., a lower synchrotron flux component). Deeper optical observations as well as refined modelling should continue to enhance our understanding of this complex source.

\end{document}